# Fibre Lasers for Gamma Colliders

L. Corner

John Adams Institute for Accelerator Science at the University of Oxford, Denys Wilkinson Building, Keble Road, Oxford OX1 3RH, UK

l.corner1@physics.ox.ac.uk

**Abstract**

The required laser parameters for recently proposed gamma-gamma collider Higgs factories are presented and reviewed in the context of new developments in high average and peak power fibre laser technology. A possible laser architecture based on fibres is proposed and some issues surrounding high average power frequency conversion to the wavelengths required for a gamma collider are discussed.

## 1. Introduction

The recent discovery of the Higgs boson has led to an interest in developing 'Higgs factories'[1], colliders specifically aimed at studying the properties of the Higgs, in addition to proposed new facilities such as the International Linear Collider (ILC)[2] or the Compact Linear Collider (CLIC)[3]. One such possible Higgs factory is a gamma-gamma collider. These have been studied in detail in the past, including proposals for a gamma-gamma or gamma/$e^-$ interaction point (IP) at the ILC and its forerunners[4,5], but recently proposals have emerged for standalone facilities dedicated to gamma collisions[6,7]. The properties and advantages of gamma colliders are outlined elsewhere[8,9]; this paper will review the properties of the lasers required for such machines and whether recent advances in laser technology mean they are now feasible propositions.

## 2. Fibre lasers for gamma colliders

Two new gamma colliders have been proposed recently, specifically SAPPHiRE[6], based on recirculating linacs, and HFiTT[7], which would be placed in the Tevatron tunnel at Fermilab. The basic principle of such a gamma collider is that Compton scattering of a laser pulse from an electron bunch (at the 'conversion point') produces high energy photons which then collide at a separate IP. To produce Higgs bosons with a mass of 126GeV would require the scattering of a laser pulse at 351nm from a beam of electrons at 80GeV, a wavelength and energy range accessible with current technology. These proposed colliders have the potential advantages over $e^+/e^-$ machines of requiring lower energy beams than are proposed for the ILC or CLIC (although of course there is a difference in the physics reach of a gamma collider and machines like the ILC), the lack of need for a positron beam which is complex to produce, and possibly a reduction in basic cost by the reuse of existing infrastructure (as in the HFiTT proposal).

It should be emphasised that these are currently very new proposals which are not completely designed or optimised. The accelerator aspects of the two proposals will not be discussed in this article, which will concentrate on the general laser requirements of these proposals or comparable machines. The laser parameters for SAPPHiRE and HFiTT are summarised in Table 1. They are broadly similar, the main difference being reflected in the repetition rates. The current published SAPPHiRE proposal has a 200kHz rep. rate, but it has been indicated that this could be reduced to 100kHz with a corresponding increase in the individual bunch charge[10]. These parameters may not represent a fully optimised final specification for such a collider, but they are unlikely to change by orders of magnitude and can therefore act as a suitable guide to the reasonable laser requirements of such colliders.

**Table 1** Laser parameters for SAPPHiRE and HFiTT gamma colliders

|  | **SAPPHiRE** | **HFiTT** |
|---|---|---|
| **wavelength** | 351nm | 351nm |
| **pulse energy** | 5J | 5J |
| **repetition rate** | 200kHz (100kHz) | 47.7kHz |
| **pulse duration** | 5ps (FWHM) | 3.5ps (FWHM) |

Some of the individual parameters are entirely reasonable given current laser technology. The energy required per pulse (~ several J), pulse durations (~ ps) and wavelength (351nm, requiring third harmonic generation of a fundamental beam at 1053nm) are all within the reach of current laser systems such as Nd:glass amplifiers. However, the average power requirements for a single pass Compton scattering system (i.e. where one laser pulse interacts with only one electron bunch) are extreme. Assuming a (slightly conservative) third harmonic generation (THG) conversion efficiency of 50%, the pulse energy requirements at the fundamental wavelength are 10J at 47.7 / 100kHz. This gives an average output power for the laser of ~ 0.5 – 1MW, something that is not possible with solid state laser technology. This also has a huge impact on the power requirement for the collider as a whole; even with an electrical – to – optical efficiency of 0.1% (better than many current systems) the electrical power needed for the laser is 500MW – 1GW. As the machine requires two laser pulses scattering from two electron bunches to create the high energy photons for collisions, this implies a power budget for the lasers alone of 1 – 2GW which is clearly impractical. It is obvious that improving the efficiency of high power laser systems is extremely important for their use in gamma colliders, but also for many other applications such as laser plasma wakefield acceleration or inertial fusion, and this is an active area of research with, for example, much effort focused on pumping the laser amplifiers with efficient diodes lasers rather than flash lamps[11].

Given these extreme demands on the laser system, the proposed solution in the past has been the use of optical cavities to enhance or recirculate one laser pulse, so that it interacts with multiple electron bunches.

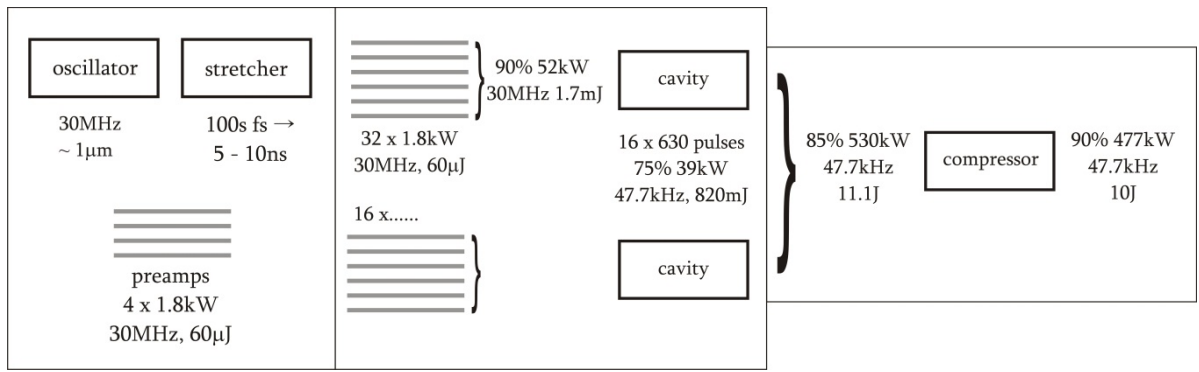

**Figure 1** Schematic of possible laser architecture for HFiTT gamma collider

This approach has been investigated for bunch trains at the ILC and its predecessor projects[12] and may be more appropriate for closely spaced bunches than for the microsecond spacing (and hence several kilometres recirculating cavity length) of the SAPPHiRE or HFiTT proposals. It is also clear that a cavity hugely increases the complexity and alignment around the conversion and interaction points, where it is also necessary to extract the electron and laser beams and have space for a particle physics detector. There are many concerns for cavity based solutions about issues such as damage, longevity and reflectivity reduction for optics exposed to a radiation hard environment, locking of the cavity length precisely to the electron bunch repetition rate and stabilisation and maintaining the spatial overlap of the laser pulse with the electron bunch[13,14]. Given these limitations and the additional complexity of an optical cavity around the conversion and interaction points, it seems that a preferable solution would be to have a single pass interaction geometry and not recycle the laser pulse, if at all possible.

As stressed before, the average power requirements for these colliders are well beyond the state of the art in laser technology. However, recent advances in the coherent combination of fibre lasers show that it might be possible to develop a laser system with both the necessary average power output and electrical efficiency to make a single pass scattering geometry for a gamma collider a feasible prospect. The ICAN project[15] has brought together expert groups who have developed proposals to combine the output of thousands of individual fibre lasers to produce pulses of ~ 30J at kHz repetition rates and pulse durations of hundreds of femtoseconds. These properties, although originally developed for driving laser wakefield acceleration experiments, are well aligned with the gamma collider requirements. Figure 1 shows a possible laser architecture for the HFiTT collider, adapted from a system proposed by Eidam et. al.[16] to fit the repetition rate requirement of HFiTT.

The proposed laser consists of an oscillator producing pulses at 1μm and 30MHz, with pulse durations of ~ 100fs. These pulses are stretched to a few nanoseconds and then split and amplified in 4 fibre pre-amplifiers, before being split again into 16 groups of 32 fibres which amplify the pulses to nearly 2mJ. The output of each group of 32 fibres is coherently combined andinjected into an enhancement cavity (see, for example, ref. 17), where the pulses are temporally stacked to create a higher energy pulse and switched out of the cavity at the accelerator repetition rate of 47.7kHz. The pulses from each cavity are coherently combined into a single pulse of 11J and then compressed to give a final laser output of 10J pulses at 47.7kHz with a duration of a few ps.

There are many challenges that have to be addressed before such a system could be built – for example, the coherent combination of very high energy pulses and the cavity switching - which are common to any application requiring this high peak and average power performance. These issues are being addressed by the groups involved in the ICAN collaboration and others worldwide. This article will concentrate on one subject specific to the gamma collider application, the frequency conversion of the fundamental at 1μm to the third harmonic at 351nm. Frequency conversion in this spectral region has been demonstrated with extremely high efficiency and peak power, for example at the National Ignition Facility (NIF)[18]. This facility has also demonstrated that it should be possible to retain the bandwidth necessary to support a pulse duration of a few picoseconds in the uv[11]. However, this performance has only been shown at very low repetition rates of less than 1Hz, and the state of the art average power handling of harmonic crystals is limited to an output of under 1kW[19], far below the ~ 250kW that would be required for conversion of the single compressed output beam of the laser shown in Figure 1. It is not clear that the required power handling capability will be developed soon (although average power handling levels for frequency conversion of 1MW have been predicted[20]), given the orders of magnitude difference in the required value over the current state of the art, although there are promising approaches using thin segmented frequency conversion crystals with gas cooling[11]. Another approach useful for the laser system proposed above would be to lower the average power handling requirement by having multiple frequency conversion points, as shown in Figure 2. Instead of a single conversion stage after the compressor, it is possible to instead convert the output of each enhancement cavity before recombination and compression. This reduces the average power at each conversion stage by a factor of 16, which is less than 20 times the state of the art, a much more approachable figure. The advanced techniques of segmentation and forced gas flow could be then used to improve the thermal management of the frequency conversion crystals and make this level of power handling a realistic proposition.

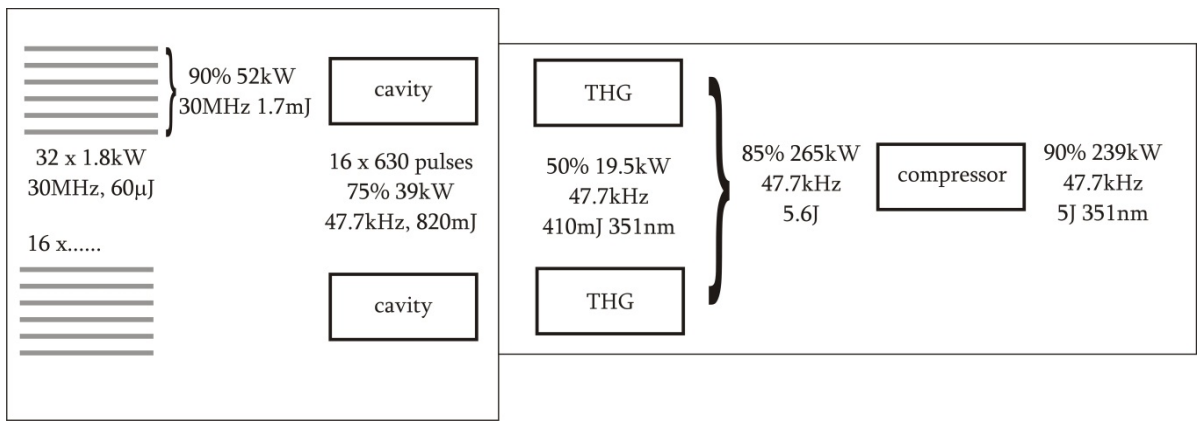

**Figure 2** Schematic of laser architecture including frequency conversion for HFiTT

## 3. Summary

The laser requirements for a gamma collider are very well aligned with those of the laser systems proposed by the ICAN/IZEST collaborations. Recent schemes for high energy, high repetition rate coherently combined fibre lasers could be straightforwardly adapted for the higher rate, lower energy requirements of the proposed gamma collider Higgs factories, thus removing the need for a complex and costly recirculation cavity around the IP of the collider.

## Acknowledgements

This work was supported by the Science and Technology Facilities Council (grant no. ST/J002011) and by the European Commission under the FP7 Research Infrastructures project EuCARD, grant agreement no. 227579. The author acknowledges helpful discussions with Dr. C. Ebbers.